# Predicting the Structural, Electronic and Magnetic Properties of Few Atomic-layer Polar Perovskite

Shaowen Xu[1,2,*], Fanhao Jia[1,2,*], Shunbo Hu[2,1], A. Sundaresan[3], Nikita Ter-Oganessian[4], A.P. Pyatakov[5], Jinrong Cheng[2,1], Jincang Zhang[2,1], Shixun Cao[1,2], and Wei Ren[1,2,†]

[1]*Physics Department, International Center for Quantum and Molecular Structures, Shanghai Key Laboratory of High Temperature Superconductors, Shanghai University, Shanghai 200444, China*
[2]*Materials Genome Institute, State Key Laboratory of Advanced Special Steel, School of Materials Science and Engineering, Shanghai University, Shanghai 200444, China*
[3]*Chemistry and Physics of Materials Unit, Jawaharlal Nehru Centre for Advanced Scientific Research, Bangalore-560064, India*
[4]*Institute of Physics, Southern Federal University, Rostov-on-Don, 344090, Russia*
[5]*M.V. Lomonosov Moscow State University, Faculty of Physics 1-2 Leninskiye Gory, GSP-1, Moscow, 119991, Russia*
[†]renwei@shu.edu.cn

*These authors contributed equally to this work.

**ABSTRACT**

Density functional theory (DFT) calculations are performed to predict the structural, electronic and magnetic properties of electrically neutral or charged few-atomic-layer (AL) oxides whose parent systems are based on polar perovskite $KTaO_3$. Their properties vary greatly with the number of ALs ($n_{AL}$) and the stoichiometric ratio. In the few-AL limit ($n_{AL} \leq 14$), the even AL (EL) systems with chemical formula $(KTaO_3)_n$ are semiconductors, while the odd AL (OL) systems with formula ($K_{n+1}Ta_nO_{3n+1}$ or $K_nTa_{n+1}O_{3n+2}$) are half-metal except for the unique $KTa_2O_5$ case which is a semiconductor due to the large Peierls distortions. After reaching certain critical thickness ($n_{AL}>14$), the EL systems show ferromagnetic surface states, while ferromagnetism disappears in the OL systems. These predictions from fundamental complexity of polar perovskite when approaching the two-dimensional (2D) limit may be helpful for interpreting experimental observations later.

**INTRODUCTION**

Few-atomic-layer transition-metal-oxide perovskites have attracted increasing interests with tremendous potential applications after the first successful synthesis down to the monolayer limit [1]. Similar to their bulk phase, a vast range of functional properties are derived from the strongly correlated states in accompany with peculiar oxygen octahedron distortions [2,3]. For example, the giant tetragonality [1], super-elasticity [4], large flexibility and piezoelectricity [5], and high-performance photodetectors [6] have quickly been demonstrated or realized on their membranes, which can be further boosted by stacking layers of such materials to create heterostructures [7,8] and by applying external strain [9-11].

Many 2D perovskite studies are based on the non-polar atomic layer (AL) oxide systems (such as $SrTiO_3$ and $BaTiO_3$) with each AL is electrically neutral, while many other polar systems such as $KTaO_3$, $LaAlO_3$, and $BiFeO_3$ etc. are also very common. These systems are inherently distinctive from non-polar systems because of the electrostatics effects. In this work, we choose $KTaO_3$ as an example to illustrate some of their most unusual characteristics. Bulk $KTaO_3$ perovskite is an incipient ferroelectric material [12] with a relatively large band gap 3.64 eV [13], which has been studied for over sixty years [14]. A number of novel properties have been discovered in this system by using methods like doping, size effect, dimension or interface engineering [15,16]. For example, people found the 2D electron gas (2DEG) with strong spin-orbit coupling on its polar surface [17,18], and the superconductivity by electrostatic carrier doping [19]. The monolayer $KTaO_3$ was predicted to be a wide band gap semiconductor [20], whereas for multilayer $KTaO_3$ it evolved from semiconductor into a state with coexistence of 2DEG and 2D hole gas (2DHG) with 16 ALs up [21] due to the so-called "polarization catastrophe" [22]. Interestingly, the 2DEG and 2DHG states could be switched to each other on top and bottom sides of the multilayer slab, by applying biaxial compressive strain.

Following the stoichiometric ratios based on the parent system of $KTaO_3$, we construct three types of few-AL KTaO systems with different chemical formulas namely $(KTaO_3)_n$, $K_{n+1}Ta_nO_{3n+1}$ to $K_nTa_{n+1}O_{3n+2}$. The schematic diagrams of their ideal structures when n = 1 is displayed in Fig. S1, and the side and top views of their structures with explicit atomic distortions shown in Fig. 1. It is easy to find that the $(KTaO_3)_n$ systems with even number of ALs (having the same number of $KO^-$ or $TaO_2^+$ layer) are electrically neutral and intrinsically symmetry-broken in $c$ direction. On the other hand, $K_{n+1}Ta_nO_{3n+1}$ and $K_nTa_{n+1}O_{3n+2}$ having odd number of ALs are electrically charged, but naturally maintain a mirror symmetry plane that is perpendicular to the $c$ axis. For convenience, we call the even number AL systems as EL, while the odd number AL systems are further divided into Type I and Type II systems.

In this work, we performed DFT simulations on three few-AL KTaO configurations. We demonstrated their stability, and identified the explicit structural distortions which exactly affect their electronic and magnetic properties. The critical thickness of various observations has been analyzed and discussed. The effects of varying the stoichiometric ratio were also compared with that from carriers doping, and the case for the non-polar $SrTiO_3$ systems.

**RESULTS AND DISCUSSION**

*The ground-state properties when n =1*

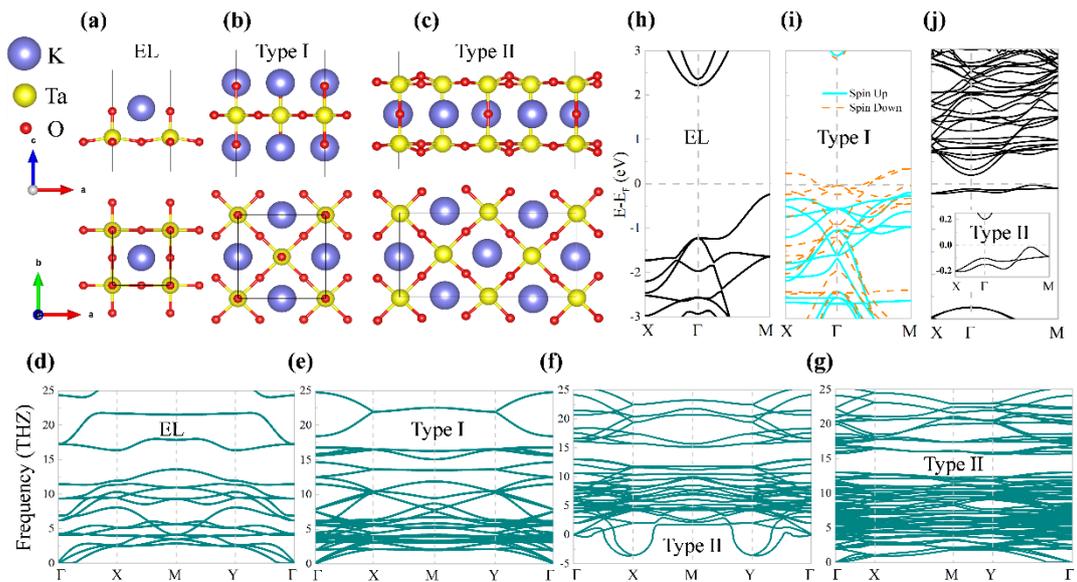

Fig. 1 **The ground-state properties of three *n* = 1 systems.** The side and top views of the relaxed structures of (a) even layer (EL) $KTaO_3$, (b) $K_2TaO_4$ (Type I), and (c) $KTa_2O_5$ (Type II). (d)-(e), The phonon spectrums of monolayer $KTaO_3$ and $K_2TaO_4$, respectively. The phonon spectrums of monolayer $KTa_2O_5$ (f) without distortions and (g) with explicit distortions. (h)-(j) The band structures of monolayer $KTaO_3$, $K_2TaO_4$ and $K_2TaO_4$, respectively. The cyan bands are spin up states, the orange bands are spin down states, and the black bands are spin unpolarized states.

The ground-state structures of three n=1 systems show different minimal periodicity as displayed in Fig. 1(a)-(c). The EL system shows the smallest in-plane periodicity, namely a 1×1 unit-cell; while the Type I system shows a √2×√2 supercell, and the Type II system requires a 2√2×√2 supercell. These ground-state structures are theoretically determined by identifying instabilities from the phonon dispersion spectra, then we accordingly lowered the symmetry and performed new structural relaxations until the instabilities were minimized. As the displayed phonon spectra in Fig. 1(d)-(g), the EL system presents the dynamic stability at the unit-cell scale with high symmetry, while the odd layer (OL) systems tend to break such high symmetry. The Type I system only maintains one mirror symmetry plane which is perpendicular to the *b*-axis. Here, we should declare that the in-plane symmetry lowing is derived by the tiny difference about 0.0001 Å of the in-plane lattice constant. However, the lattice instability of high-symmetry Type II system is dramatic as displayed in Fig. 1(f). As a result, the significant out-of-plane distortions are necessary, and a periodicity twice along *a*-axis is reserved. In order to further eliminate the concern of structural instability, we have examined their elastic and thermodynamic stabilities by calculating their elastic constants (see Table S1 for the data and the details of the criteria) and performing *ab* initio molecular dynamics (AIMD) (see Fig. S2).

We display the PBE+*U* band structures in Fig. 1(h)-(j), whose corresponding layer resolved and atomic-orbital projected density of states (PDOS) are shown in Fig. S3. It is easy to find that the EL system and Type II system are semiconductors, while the Type I is a spin-polarized half metal. The calculated band gap of EL system is 3.51 eV by using HSE06 functional, which is well consistent with previous report [20]. This band gap is comparable with the value of the bulk; however, we could not directly conclude that the quantum confinement has almost no effect in this system because the dependence on the thickness is significant as $n_{AL}$ increases. Similar to the bulk, the EL system has the valence band maximum (VBM) dominated by the O-$p$ orbitals with very slight hybridization of Ta-$d$ orbitals, while the conduction band minimum (CBM) is mainly derived from Ta-$d$ orbitals. However, there are two major differences between bulk and monolayer: one is that the VBM of the monolayer only has the contributions from $TaO_2$ ALs with the contributions of KO ALs shifting about 1.3 eV below the VBM; the other one is that the KO ALs only donate $p_z$ component nearby the Fermi level while the $TaO_2$ ALs only donate $p_x$ and $p_y$ components (Fig. S3).

The Type I system has an additional KO AL on the top of EL system. Due to this extra charged layer, the system is metallic and shows the feature of lacking electrons. Moreover, this system is spin polarized with a magnetic moment of 1 $\mu_B$/u.c., which is derived from the O-$p$ orbitals of both the KO AL and $TaO_2$ AL. Then the situation changes again when turning to the Type II system. The Type II system adds an additional $TaO_2$ AL on the top of EL system, which should be a metallic system due to the excess electrons. However, we find that only its high symmetry phase is metallic, which seems unstable indicated by the phonon calculation. After including the explicit atomic distortions, the system is stabilized, accompanied with a Peierls-like phase transition to a semiconductor phase with a band gap about 0.36 eV. The partially occupied Ta-$d$ orbitals are split into two fully occupied bands, and the rest bands are empty. These two occupied Ta-$d$ bands correspond to the two excess electrons of the additional $TaO_2$ AL, which form flat bands in very narrow energy window. It is worth mentioning that the flat bands are recently found to be highly relevant with the unconventional superconductivity in the twistronics [23] and other strongly correlated properties [24].

### *The strong dependence on layer thickness*

Unlike the conventional van der Waals 2D materials, the layers in few-AL perovskite are chemically stacked by the ionic or stronger bonds. As a result, there is no complex issues of diverse stacking [25] and twisting angle [26] problems. However, the physical properties still vary greatly when approaching the 2D limit, and the situation is further more complicated if we consider polar materials. The optimized in-plane lattice parameters are listed in Table S2 and Table S3, which slightly increase as the thickness increases. In Fig. S4, we displayed the thickness dependence of the out-of-plane polarization $P_z$ of EL KTaO and SrTiO systems. The latter was taken as a reference, whose magnitude of $P_z$ decreases monotonously as the number of AL increases, which could be attributed to the gradually weakened surface effect. The EL KTaO systems behave similarly if the $n_{AL} \leq 14$, but show an anomalous upturn of $P_z$ after the $n_{AL}$ reaching 16. This kink is due to the collective electrons and holes on the top and bottom surfaces are further separated as the thickness of the slab increases.

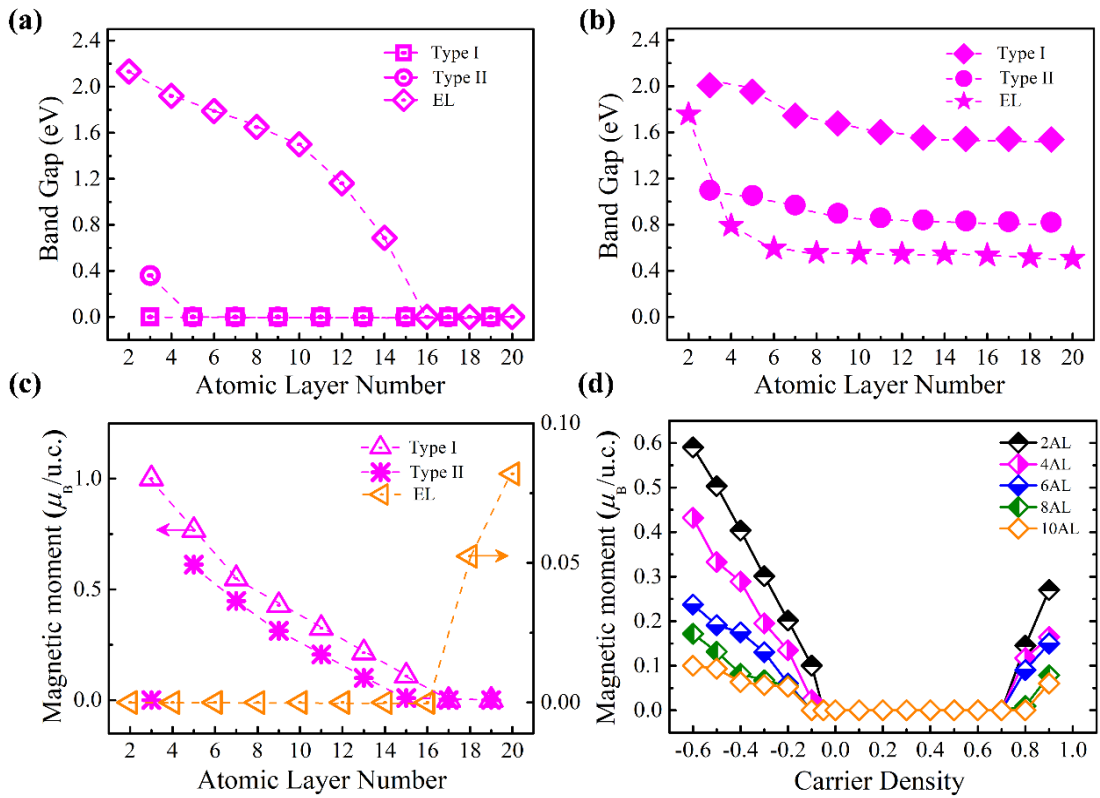

Fig. 2 **The dependence of electronic and magnetic properties on the layer thickness.** The band gaps as functions of AL thickness for three configurations of (a) KTaO systems and (b) SrTiO$_3$ (SrTiO) systems. (c) The magnetic moments ($\mu_B$/u.c.) as functions of AL thickness, here the "u.c." is defined by the in-plane area of EL unit-cell displayed in Fig. 1. (d) The effect of doping carriers on the magnetic moment of few EL systems. The carrier density is ranged from -0.1 to -0.6 corresponding to electron

concentration $n \sim 5\times10^{12}$ to $8\times10^{13}$ e/cm$^2$.

We displayed the band gaps as functions of AL thickness of three-type configurations of few-AL KTaO systems and SrTiO systems in Fig. 2(a) and 2(b), respectively. It is easy to find that the few-EL systems of both compounds show their largest band gap when they approach the monolayer limit, which could be accounted by the quantum confinement effects [27,28]. On the other hand, the OL systems are always metallic except for the Type II (n=1) $KTa_2O_5$, while all SrTiO systems are semiconductor and their band gaps tend to converge when the AL thickness is larger than 8. Remarkably, we notice that the few-EL KTaO systems are metallic when the AL thickness is larger than 14, which is consistent with previous report [21]. This could be explained by the "polar catastrophe" of perovskite materials when an electrostatic instability happens as layer thickness is accumulated to certain threshold. Further, there are many compensation mechanisms to stabilize such as electrostatic shielding by 2DEG, charge transfer, surface reconstruction, chemical doping and adsorption [15-17,19,29].

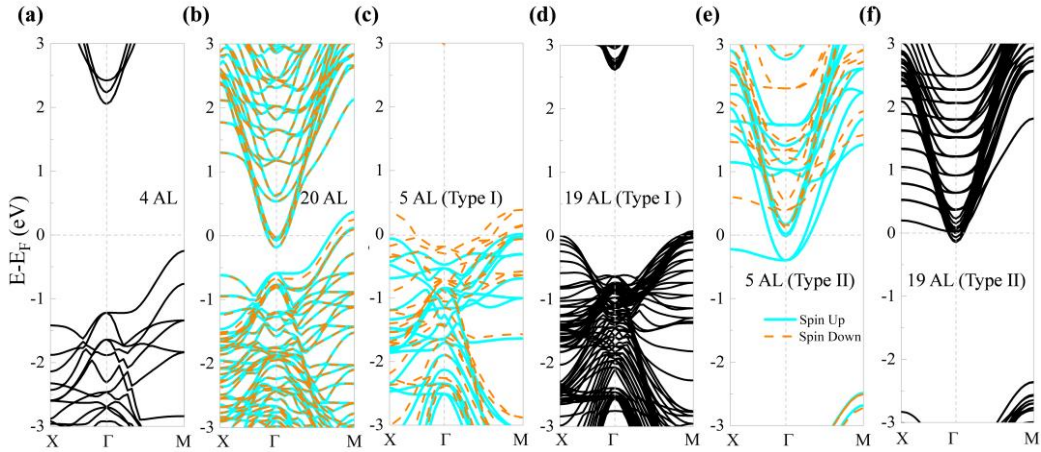

Fig. 3 **The band structures for the AL thickness close to the values 4 and 20.** (a)-(b) are for 4AL and 20AL EL; (c)-(d) are the Type I; and (e)-(f) are the Type II systems.

As displayed in Fig. 2(c), in accompany with the insulator-metal transition, there is a nonmagnetic - ferromagnetic phase transition, and the relatively weak magnetic moments start to increase as the number of EL increases. According to the spin polarized band structure of 20 AL system in Fig. 3(b), the magnetic moment of EL system is mainly from the hole pocket derived from the surface states in the KO termination. The mechanism of the ferromagnetism can be further explained by using the Stoner model [30]. On the other hand, apart from the $KTa_2O_5$ system, the OL systems are half-metal in the few-AL limit. The magnetic moment strongly depends on the density of states (DOS) near the Fermi level, thereby when increasing the density of carriers, the corresponding magnetic moment increases (shown in Fig. 2(d)). Another interesting fact is that doping electron carriers to the EL system is more easily to be spin polarized than doping hole carriers. Moreover, we also notice that the Type I systems with hole pockets always show slightly larger magnetic moments than the Type II systems given the same thickness. On the contrary, the magnetic moments of odd AL systems gradually decrease as the number of AL increases, and disappear after the number of AL reaching 17. By comparing the band structures of 5 AL in Fig. 3(c) and 3(e), and 19 AL in Fig. 3(d) and 3(f), we find that eventually the 19 AL systems show much smaller Fermi pockets than the 5 AL systems, which explains the elimination of ferromagnetism in the very thick OL systems.

*Layer-resolved properties*

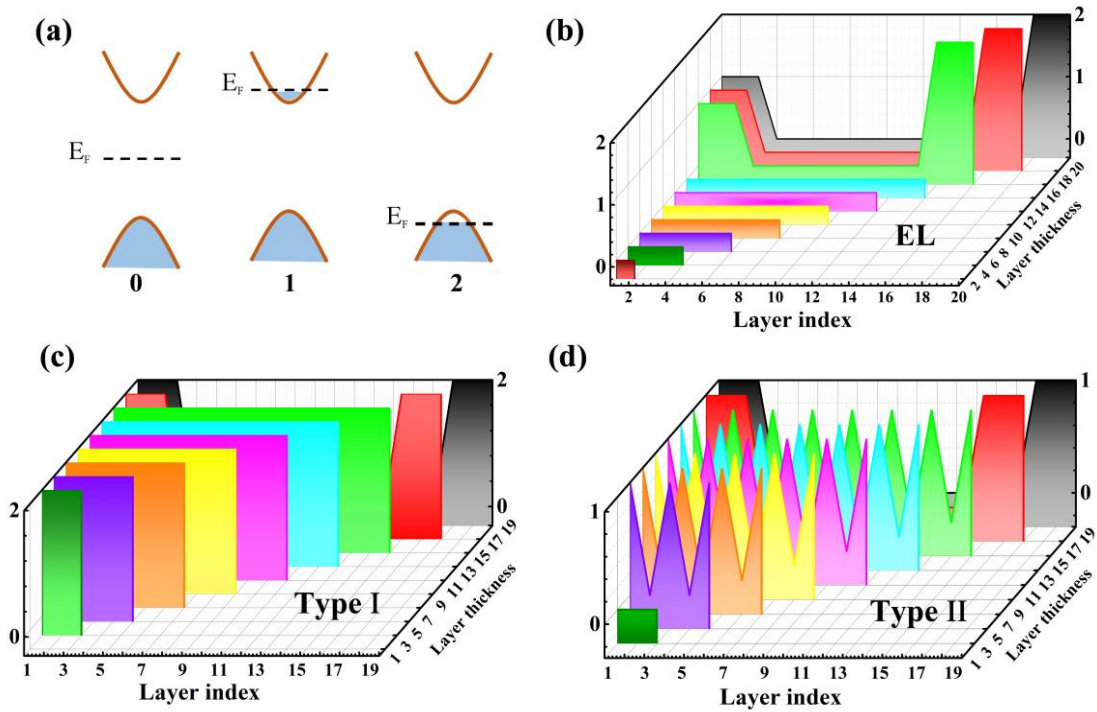

Fig. 4 **The layer-resolved phase diagram to illustrate the critical thickness.** (a) The definition of three phases: 0 represents the AL is insulator; 1 represents the AL with electron pocket; 2 represents the AL with hole pocket. The dashed line represents the position of Fermi level, and the shaded parts represent the occupied electronic states. The phase diagram of (b) EL systems, (c) Type I systems, and (d) Type II systems, where the tick of *x*-axis is the index of resolved layer, the tick of *y*-axis is the layer thickness, and the tick of *z*-axis is the phase index defined in (a).

To better understand the critical thickness and the intriguing results of KTaO systems, we studied their layer-resolved properties, and presented them in the phase diagram in Fig. 4(a), where each AL can be labeled by one of three types of phases, namely the insulator phase 0, the metal phase 1 with electron pocket, or the metal phase 2 with hole pocket. As seen in Fig.4(b), the EL systems are insulator when $n_{AL} \leq 14$. But when $n_{AL} \geq 16$, they transform into a phase where 2DEG and 2DHG exist simultaneously at two opposite sides of the slab. The 2DEG is localized at the surface with $TaO_2$ AL as termination, while the 2DHG is localized at the surface with KO AL as termination. From Fig.4(c), the Type I systems with $n_{AL} \leq 15$ are all metallic phase 2. After $n_{AL} \geq 17$, only the surface ALs of the slabs remain metallic state with the feature of hole pockets, while leaving the central ALs insulator phase 0. Type II systems are more complex as shown in Fig.4(d). When $n_{AL}= 3$, the ALs are in insulator. When $3< n_{AL} \leq 15$, the ALs have both insulating (phase 0) and metallic (phase 1) states. More specifically, all TaO ALs are metallic with electron pockets, while all KO ALs are in insulator phase. Then, after $n_{AL} \geq 17$, only the surface ALs remain in metal phase with the feature of electron pockets, leaving all those central ALs in insulator phase. It is extremely interesting that all three types of our studied KTaO systems show a critical thickness about 16 ALs, above which the surface states emerge on the two sides of the slab models. Such surface states generally have common thickness and take place at three outmost ALs, as illustrated in Fig.4(b)-(d).

In summary, we predicted some fundamental properties of few-AL systems based on the polar perovskite $KTaO_3$ when approaching the 2D regime. In the few-AL limit, the EL systems are semiconductors whose band gaps gradually decrease as the $n_{AL}$ increases; the OL systems are spin polarized metallic except for the unique $KTa_2O_5$ case which is a semiconductor due to the large Peierls distortions. When $n_{AL} > 14$, the EL system transforms to a spin polarized metal phase with the coexistence of 2DEG and 2DHG on opposite surfaces, while the OL systems transform to a spin unpolarized metal phase. All surface states of three studied configurations are localized within a critical thickness about 3AL. Our results are expected to stimulate the synthesis of few-AL polar perovskite materials, as well as help design electronic and spintronic devices based on the conductive oxide interfaces [31].

**Methods**

First-principles calculations were performed within the generalized gradient approximation (GGA) in the form proposed by Perdew, Burke, and Ernzerhof (PBE) [32], which is implemented in the Vienna *ab initio* simulation package (VASP) [33]. The energy cutoff was chosen to be 600 eV. An additional effective Hubbard $U_{eff}$ = 3.0 eV for 5$d$ states of Ta was used to reduce the delocalization error [21]. The Γ centered k-grids were adopted with the smallest allowed spacing ~ 0.04 /Å$^{-1}$ and ~0.02 /Å$^{-1}$ for semiconductor phase and metal phase, respectively. The atomic positions were fully relaxed until the maximum force on each atom was less than $10^{-3}$ eV/Å. To address the well-known problem of underestimating the band gap of PBE, the screened hybrid functional HSE06 [34] was applied to calculate the band gaps. A vacuum layer more than 20 Å was adopted to avoid the interaction between periodic images. The phonon dispersion was calculated by density-functional perturbation theory (DFPT) as implemented in the PHONOPY package [35,36]. The carrier doping was done by changing the total number of electrons of system, and a compensating jellium background of opposite charge was added to maintain charge neutrality.

**ACKNOWLEDGMENTS**

This work was supported by the National Natural Science Foundation of China (Nos. 51861145315, 51911530124, 51672171, 12074241), Shanghai Municipal Science and Technology Commission Program (Nos.19010500500 and 20501130600), State Key Laboratory of Solidification Processing in NWPU (SKLSP201703), Austrian Research Promotion Agency (FFG, Grant No. 870024, project acronym MagnifiSens), and

Independent Research Project of State Key Laboratory of Advanced Special Steel and Shanghai Key Laboratory of Advanced Ferrometallurgy at Shanghai University. F.J. is grateful for the support from the China Scholarship Council (CSC). APP and NVT acknowledge financial support by the Russian Foundation for Basic Research grant No. 18-52-80028 (BRICS STI Framework Programme).